\documentstyle[epsf,prd,aps,floats]{revtex}
\input{psfig.tex}

\tighten
\begin{document} 
\draft
\preprint{Imperial/TP/96-97/56 
}
\twocolumn[\hsize\textwidth\columnwidth\hsize\csname @twocolumnfalse\endcsname
\title{Non-Gaussian sampling effects on the CMB 
power spectrum estimation}
\author{Charlotte Cheung and Jo\~ao Magueijo}
\address{Theoretical Physics, The Blackett Laboratory,
Imperial College, Prince Consort Road, London, SW7 2BZ, U.K.}
\maketitle
\begin{abstract}
We introduce a non-Gaussian model of structure formation, and show
how it produces surprising sampling effects on the estimation of
the CMB power spectrum. Estimates of the average
power spectrum based on small sky
patches produce qualitatively different results from patch to 
patch, and vary also with the patch size. For instance a 
system of secondary Doppler peaks is observed
in small enough patches, but at different positions in different patches.
However, no secondary Doppler peaks are observed in large sky patches.
These effects question usual methods of power spectrum estimation,
and techniques like bootstrapping.
The non-Gaussianity of this model fails to be detected by 
standard tests.  
\end{abstract}

\date{\today}

]

\renewcommand{\thefootnote}{\arabic{footnote}}
\setcounter{footnote}{0}

Considerable effort is currently being put into the estimation of the
power spectrum $C_\ell$ of CMB temperature fluctuations 
\cite{gorski,jaffe,teg}.
The $C_\ell$ are expected to provide a wealth of information on theories of
the early Universe, discriminating between paradigms \cite{def,infl}, 
and allowing highly accurate measurements of cosmological parameters 
\cite{hobs,bet}. 
The assumption that the temperature
fluctuations constitute a Gaussian random field plays a central role in
nearly all data-analysis methods (see however \cite{mem}). 
This assumption has been checked in some
data sets (eg. \cite{kogut}), although one can argue that the tests
applied may not be conclusive \cite{fermag}. One may certainly
devise subtle non-Gaussian theories which pass the 
traditional Gaussianity tests. 

Although non-Gaussianity cannot show up at the level of the average 
power spectrum $C_\ell$, it may affect its estimation from a single
sky. Larger $C_\ell$ cosmic variance error bars, and correlations between 
all-sky $C_\ell$ are familiar non-Gaussian effects known to be present 
in some models \cite{ngsp}. These effects would certainly affect 
comparison between theoretical and experimental $C_\ell$. 
Correlations between $C_\ell$ would allow more structure to be 
present in each observed $C_\ell$ spectrum than in the average $C_\ell$. 
Also one would have to revise the allocation of error bars 
to experimental $C_\ell$ estimates, since these errorbars normally 
include cosmic  variance.

More interesting still is the possibility that non-Gaussianity might
affect in a similar way 
sampling statistics, that is, cosmic variance in sky patches
and the coordination of estimates derived from different sky patches.
It could for instance happen that estimates of the 
{\it average} power spectrum based 
on small sky patches produce different results 
from patch to patch. New results again could be obtained by 
changing the size of the patch. 

To give a motivated example, 
take a defect theory known not to display 
secondary Doppler peaks \cite{andy}. It could happen that power spectrum
estimates using small patches of the sky revealed a structure of 
secondary Doppler peaks, but that different patches saw peaks 
at different positions. The $C_\ell$ average could then 
display no secondary Doppler peaks, and it is
this average that would be sampled by a large enough patch of the sky.
We will see that this non-Gaussian effect may be achieved with
a coherent \cite{us} non-Gaussian source. 
For large sky patches such a source mimics the 
incoherent phenomena studied in \cite{andy}.
There is therefore a loophole in the argument
connecting absence of secondary peaks and incoherence, if one
allows for non-Gaussianity.

As we shall see, it is possible to construct
theories which display this effect while still 
passing the traditional Gaussian tests.
This is somewhat disturbing, as for signals coming from
such theories the usual methods for estimating the power 
spectrum would be grossly wrong, and in any case miss the theory's
rich structure of spectra. Techniques like bootstrapping 
would not be appropriate. 

Here we present a simple model with all the properties 
mentioned above. 
We regard it as a valid model of structure formation 
deserving attention in its
own right. However we reserve for a longer publication \cite{charlo2}
a presentation of the details of the model, and concentrate
here on its unusual sampling effects.
The idea is that structure is due to seeds in the early Universe, which
act as a separate inhomogeneous component. All we know about them is that
their formation and 
evolution satisfies energy conservation and that their power spectrum
respects the causality constraints \cite{traschen}. 
In this sense these seeds 
are like the seeds in the isocurvature models of \cite{peebles}. 
They represent
an ad-hoc construction led merely by the guiding principles of causality
and energy conservation.

However we postulate some dynamics. We require these seeds 
to be explosions, as in the mimic models discussed in \cite{neil,hu}. 
They differ from mimic explosions only in that they do not explode
at the Big Bang, but at a given blast time $\eta_b$. When they explode
they start seeding structure. Before the explosion they merely lurk about
imparting no gravitational action on the surrounding matter. 
As in \cite{neil,hu} we require these seeds to 
be scale-invariant.

A final but central ingredient of the model is that the blast time
$\eta_b$ is random, with a probability distribution  $P(\eta_b)$,
and is uncorrelated beyond a given distance $\xi$. 
This is the
only qualitative novelty of our model, but this feature will be responsible
for all the weird non-Gaussian behaviour to be derived. 
In summary 
structure is seeded by bombs which have always been around, 
but which remain dormant for part of their lives,
then explode, not at a pre-set time, but whenever it pleases
them. Do not confuse this scenario with the explosive scenarios proposed 
in \cite{ostriker}. Our explosions satisfy energy conservation separately,
and only interact with the rest of the Universe gravitationally.

More concretely we define our bombs by means of their stresses. 
If $\Theta_{\mu\nu}$ is their stress energy tensor, then we have 
$\Theta_{ij}=p^s\delta_{ij}+
(\partial_i\partial_j-1/3\delta_{ij}\partial^2)\Theta^s$. 
The scalars $p^s$ and $\Theta^s$ may then be Fourier transformed.
We choose $\Theta^s=0$ and $p^s=0$ for $\eta<\eta_b$, but
\begin{equation}
  p^s=
    {1\over \eta^{1/2}}{\sin{Ak(\eta-\eta_b)}\over Ak(\eta-\eta_b)}
\end{equation}
for $\eta>\eta_b$. For definiteness we take $A=1$.
Switching on stresses rather than energy ensures that we do 
not violate energy conservation. It is also
reminiscent of the switching
on of topological defects during a phase transition \cite{joaoasakid}.
The other components of the stress energy tensor are  determined
by the energy conservation equations. 

The above specifies each individual explosion. If we set up a Poisson
process of such explosions (with fixed $\eta_b$) 
then the resulting field power spectrum equals the
density of explosions times the square of the Fourier transform
of each individual explosion \cite{charlo2}.   
If furthermore the density of explosions 
is very large we have a Gaussian random field within
each correlated region. We assume that the
source is coherent \cite{us} within each correlated region.
The calculation of the $C_\ell$ conditional to a given
$\eta_b$ can then be made in the usual way \cite{neil,hu}.

We finally choose a distribution of blast times lifted from radioactive
decay:
\begin{equation}
P(\eta_b)={e^{-\eta_b/\tau_b}\over \tau_b}
\end{equation}
where $\tau_b$ is the average blast time. 
We assume that the spatial correlation between the blast times 
$\eta_b$ dies off at a scale
$\xi$ which when projected on the last scattering surface obeys
$\xi\ll 4\pi$ (in radians).

\begin{figure}
\centerline{\psfig{file=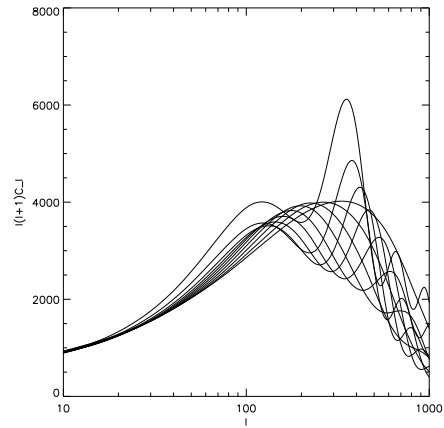,width=6 cm}}
\caption{The spectrum of angular power spectra $C_l(\eta_b)$ which would 
be sampled in small sky patches. We only show spectra with 
a non-negligible probability, for $\tau_b<\eta_*$, where $\eta_b$
is the conformal time at last scattering.}
\label{fig1}
\end{figure}

The CMB anisotropies produced by such a theory exhibit
a rich structure of power spectra. For
patches of sky with a size $L$ smaller than $\xi$ the system is best
described by noting that the temperature probability distribution 
conditional to a given $\eta_b$ is Gaussian. The observed
power spectrum ${\hat C_\ell}$ conditional to $\eta_b$ is the usual
$\chi^2$ distribution. We define the conditional average power spectrum as:
\begin{equation}
C_\ell(\eta_b)=\int d{\hat C_\ell} {\hat C_\ell} P({\hat C_\ell}|\eta_b)
\end{equation}
The sample variance is the usual
\begin{equation}
\sigma^2({\hat C_\ell})={2\over N_\ell }C^2_\ell(\eta_b)
\end{equation}
where $N_\ell=(2\ell+1)f_{\rm sky}$ is the total number 
of modes contributing to the estimate, with $f_{\rm sky}$ 
the fraction of sky covered by the patch.
Conditional to a given $\eta_b$ there are no correlations between
the ${\hat C_\ell}$ (other than the ones mentioned
in \cite{hobsmag}, which we assume have been bypassed).
This is the simplest description for patches with $L<\xi$: 
``a spectrum of spectra'' $C_\ell(\eta_b)$, and the distribution 
of the parameter $\eta_b$. In other words, the statistical 
ensemble is best described 
in terms of a set of Gaussian sub-ensembles, with a power spectrum indexed
by a variable $\eta_b$. From the point of view of the overall ensemble
$\eta_b$ is itself a random variable, with a distribution $P(\eta_b)$.
In Fig.~\ref{fig1} we show the various power spectra $C_l(\eta_b)$ which
any estimate derived from small patches would produce.

Applying the traditional description,
in terms of an average $C_\ell$ plus a 
sample variance error bar $\sigma^2({\hat C_\ell})$, is 
a bad idea for patches with $L<\xi$. Although the temperature 
distribution conditional to $\eta_b$ is Gaussian, the marginal 
distribution is not. Non-Gaussian effects on the power spectrum therefore
emerge. The marginal average power spectrum is
\begin{equation}
C_\ell={\langle {\hat C_\ell} \rangle}=
\int d{\hat C_\ell} d\eta_b {\hat C_\ell} P({\hat C_\ell},\eta_b)
\end{equation}
Using $P({\hat C_\ell},\eta_b)=P({\hat C_\ell}|\eta_b)P(\eta_b)$
leads to
\begin{equation}
C_\ell=\int d\eta_b P(\eta_b) C_\ell(\eta_b)
\end{equation}
The sample variance is now 
\begin{equation}
\sigma^2({\hat C_\ell})={\left(1+{2\over N_\ell}\right)}
\int d\eta_b C^2_\ell(\eta_b)P(\eta_b) -C_\ell^2
\end{equation}
always larger than the Gaussian sample variance $2C_\ell^2/N_\ell$
by virtue of the Schwarz  inequality. 
If the distribution $P(\eta_b)$ is reasonably peaked and
$C_\ell(\eta_b)C_\ell^{''}(\eta_b)\ll C_\ell^{'2}(\eta_b)$
we get the ``error propagation formula'':
\begin{equation}
\sigma^2({\hat C_\ell})={2\over N_\ell}C_\ell^2 +{\left({2\over N_\ell}
+1 \right)} 
{\left(\partial C_\ell\over \partial \eta_b\right)}^2
\sigma^2(\eta_b)
\end{equation}
Even if $N_\ell=\infty$, there would be a residual, purely
parametric, cosmic variance:
\begin{equation}
\sigma^2({\hat C_\ell})=
{\left(\partial C_\ell\over \partial \eta_b\right)}^2
\sigma^2(\eta_b)
\end{equation}
There are also correlations between ${\hat C_\ell}$:
\begin{equation}
{\rm cov}({\hat C_\ell},{\hat C_{\ell '}})=
\int d\eta_b C_\ell(\eta_b)C_{\ell '}(\eta_b) P(\eta_b)- C_\ell C_{\ell'}
\end{equation}
Larger sample variance error bars and correlations between $C_\ell$
(``cosmic covariance''  \cite{ngsp}) no doubt complicate 
comparison between theory and experiment. One is better off without
them, which in this case means abandoning the $C_\ell$ description,
and taking up instead $C_\ell(\eta_b)$ and $P(\eta_b)$.

\begin{figure}
\centerline{\psfig{file=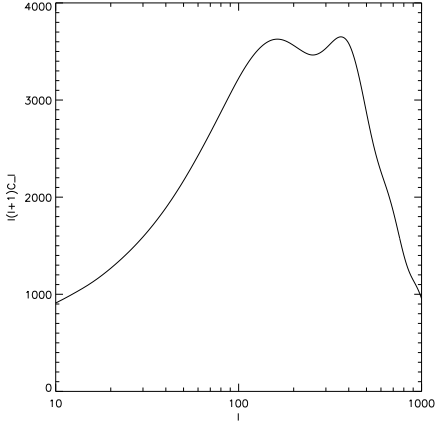,width=6 cm}}
\caption{The marginal average power spectrum $C_\ell$. This is
a bad description for observations on small patches with $L<\xi$.
It does however represents the power spectrum sampled 
by large patches.}
\label{fig2}
\end{figure}
In Fig.~\ref{fig2} we plotted the average power spectrum $C_\ell$.
Clearly the larger sample variance error bars merely hint at the fact
that the average power spectrum $C_\ell$ fails to describe what would
be observed in each small patch of the sky with $L<\xi$. 
What is seen in any patch is one of the $C_\ell(\eta_b)$ spectra 
with Gaussian fluctuations about it,
not the $C_\ell$ spectrum with a larger random scattering about it.

If we choose a patch with $L\gg\xi$ then the situation is different.
Power spectrum estimates based on the whole patch amount to an average
of the estimates that would be supplied by each correlated region with
size $\xi$. 
Let us index these regions by $i$ and let there be $N=(L/\xi)^2$ 
of them, so that 
${\hat C_\ell}=\sum {\hat C^i_\ell}/N$.
The statistics of ${\hat C^i_\ell}$ are the same as above 
with $N_\ell=NN^i_\ell$, and so
we have 
\begin{equation}
\sigma^2({\hat C_\ell})={\left({1\over N}+{2\over N_\ell}\right)}
\int d\eta_b C^2_\ell(\eta_b)P(\eta_b) -{C_\ell^2\over N}
\end{equation}
and
\begin{equation}
{\rm cov}({\hat C_\ell},{\hat C_{\ell '}})={1\over N}{\left(
\int d\eta_b C_\ell(\eta_b)C_{\ell '}(\eta_b) P(\eta_b)- C_\ell C_{\ell'}
\right)}
\end{equation} 
If the distribution $P(\eta_b)$ is reasonably peaked we get 
\begin{equation}
\sigma^2({\hat C_\ell})={2\over N_\ell}C_\ell^2 +{\left({2\over N_\ell}
+{1\over N}\right)} {\left(\partial C_\ell\over \partial \eta_b\right)}^2
\sigma^2(\eta_b)
\end{equation}
We see that the excess sample variance, and the correlations between
${\hat C_\ell}$ get smaller as the number of uncorrelated regions in the patch
$N$ increases. As $N\rightarrow\infty$ the residual parametric cosmic
variance mentioned above also disappears.
Essentially what happened is that for large patches 
the average power spectrum $C_\ell$ has become
representative of the observed ${\hat C_\ell}$. 
Observed ${\hat C_\ell}$ for large patches do scatter
around a curve as in Fig.~\ref{fig2}. 
\begin{figure}
\centerline{\psfig{file=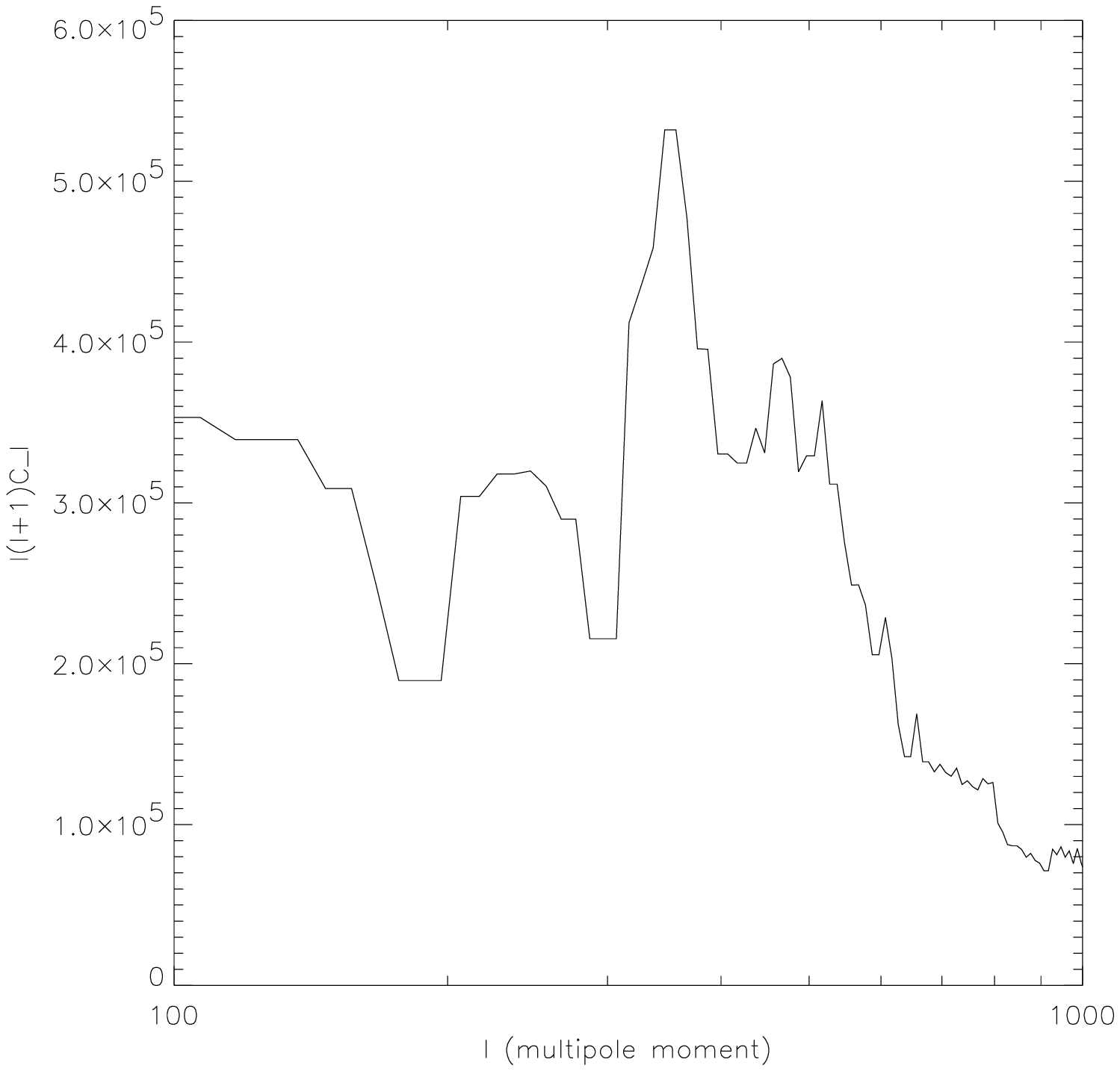,width=4 cm}}
\centerline{\psfig{file=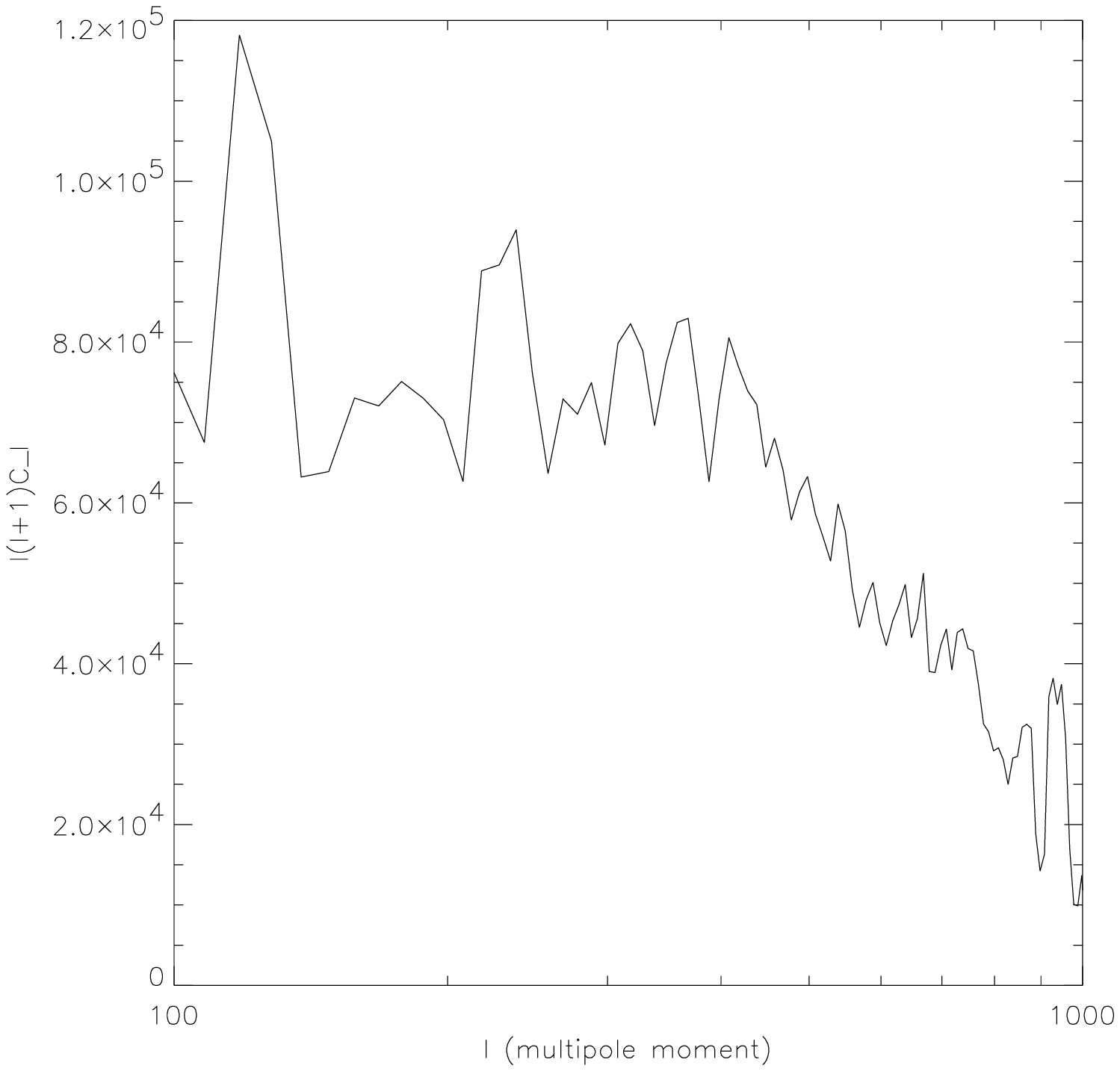,width=4 cm}}
\caption{Measured power spectra in a patch $4^\circ$ across,
and in a patch $20^\circ$ across. A maximum likelihood estimate
of the average power spectrum based on small patches would
reveal a system of secondary peaks, at different positions in 
different patches. The same exercise in a large patch would
suggest absence of secondary peaks.}
\label{fig3}
\end{figure}

We illustrate this phenomenon further in Fig.~\ref{fig3}.
We chose $\xi=5^\circ$, generated maps \cite{charlo2}, and found 
the {\it measured} power spectrum in a square patch $4^\circ$ across,
and in a patch with $20^\circ$.  
Secondary Doppler peaks are observed in each realization for
small sky patches. However the peak's positions vary from
patch to patch, or from realization to realization. Hence the
average over realizations does not exhibit secondary Doppler
peaks. Similarly estimates based on large patches do not show 
secondary Doppler peaks. This effect was found by Turok
in cosmic string simulations \cite{neilmaps}. 

\begin{figure}
\centerline{\psfig{file=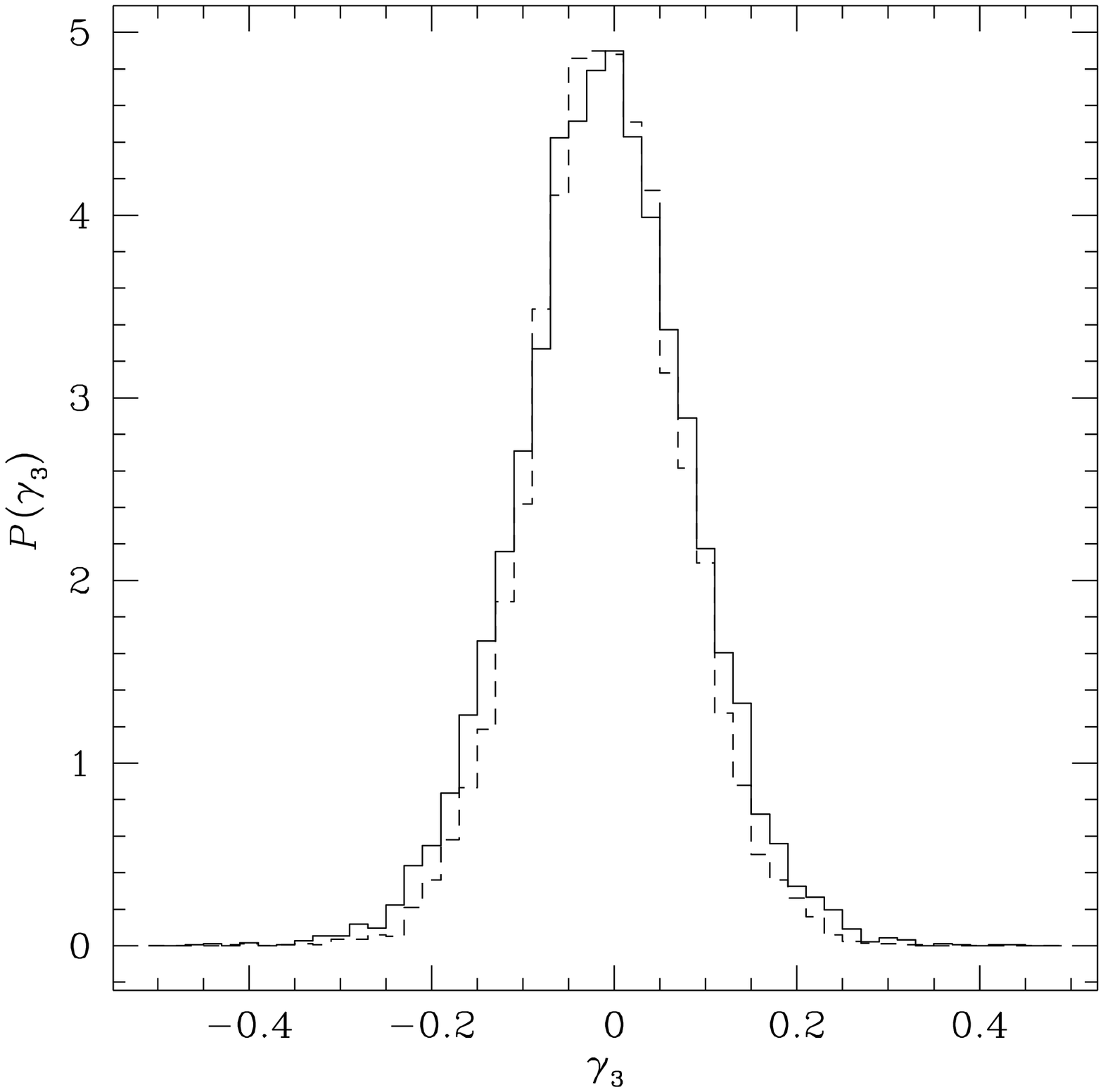,width=6 cm}}
\centerline{\psfig{file=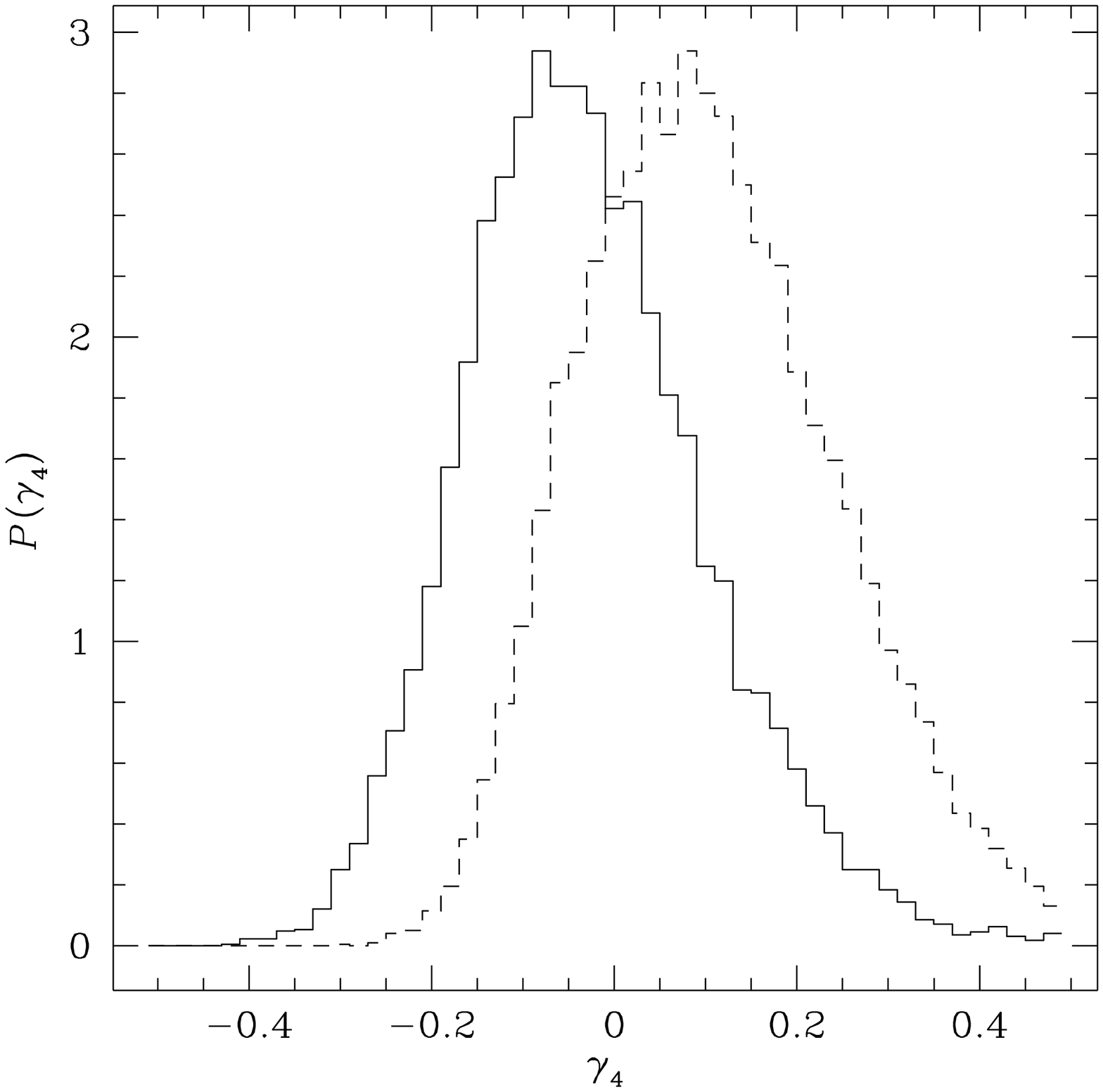,width=6 cm}}
\caption{Histograms of skewness $\gamma_3$ and kurtosis 
$\gamma_4$ as estimated from 25600 pixels for a Gaussian
theory (line) and our theory (dash). These histograms have 
used 10000 realizations.}
\label{fig4}
\end{figure}

The theory we have just proposed is somewhat distressing.  
The non-Gaussianity of its maps
fails to be detected by the traditional tests, such as skewness 
and kurtosis,
density of peaks above a given height, genus number and the 
3-point correlation
function. We show how this is the case with the skewness $\gamma_3$
and the kurtosis $\gamma_4$. 
In Fig.\ref{fig4} we plot histograms of $\gamma_3$ and $\gamma_4$
as estimated from 25600 pixels for 10000 realizations. Plots
with dashes correspond to our theory, with lines to 
a Gaussian theory with the same average $C_\ell$ spectrum. 
As far as skewness is concerned
the two theories are essentially the same. Our theory has a slightly
positive kurtosis, but not beyond what could be explained from
cosmic variance for a Gaussian theory. 
Even this latter effect can be eliminated by changing
the distribution $P(\eta_b)$ so as to ensure that the most likely
power spectra $C_\ell(\eta_b)$ have the same
$\int d\ell \ell^2 C_\ell$. This amounts to reducing the 
probability of patches with $\eta_b\approx 0$.  

Despite its apparent Gaussianity, the theory proposed shows 
a sampling effect on the power spectrum that invalidates traditional 
estimation methods. In particular bootstrapping becomes very misleading.
One should rather aim at independent power spectrum estimates 
for all possible combinations of patch location and size. 
The outcome would be $C_\ell (\eta_b)$, $P(\eta_b)$, and 
$\xi$. A standard estimation of the power spectrum would
be highly misrepresentative of the actual structure of spectra
of the theory.

There is an irritating inconsistency between power spectrum estimates
produced by different experiments, using different patches of the sky,
or patches with different sizes \cite{ngsamp}. Although these may be due 
to imperfect foreground subtraction, or calibration errors,
it is curious to point out that 
such inconsistencies might not be inconsistencies at all. 
They may be telling
us about the unusual power spectrum structure non-Gaussianity may induce. 

Finally our theory shows that maybe we should be more careful
when ruling out defects on the grounds of unflattering average power
spectra $C_\ell$ \cite{ne,she,abr}. We know very little about non-Gaussianity
in these models, but nothing tells us that comparison between 
theoretical and experimental $C_\ell$ should be made 
in the same way as for 
Gaussian theories.

ACKNOWLEDGEMENTS: We would like to thank Kim Baskerville and 
Pedro Ferreira for many
helpful comments. We have used a modified version of the Boltzmann
code of Seljak and Zaldarriaga \cite{selz} in our calculations.
C.C was supported by PPARC and J.M. by the Royal Society.

\end{document}